\documentclass[twocolumn,showpacs,prb,amsmath,amssymb]{revtex4}
\usepackage{graphicx}
\usepackage{array}

\begin{document}

\title{Excitons in core-only, core-shell and core-crown CdSe nanoplatelets:
interplay between in-plane electron-hole correlation, spatial and dielectric confinement}
\author{Fernando Rajadell, Juan I. Climente, Josep Planelles}
\email{josep.planelles@uji.es}
\affiliation{Dept. de Qu\'imica F\'isica i Anal\'itica, Universitat Jaume I, 12080, Castell\'o, Spain}
\date{\today}

\begin{abstract}
Using semi-analytical models we calculate the energy, effective Bohr radius and radiative 
lifetime of neutral excitons confined in CdSe colloidal nanoplatelets (NPLs).
The excitonic properties are largely governed by the electron-hole in-plane correlation,
which in NPLs is enhanced by the quasi-two-dimensional motion and the dielectric mismatch with
the organic environment.
In NPLs with lateral size $L \gtrsim 20$ nm the exciton behavior is essentially that in a quantum well, 
with superradiance leading to exciton lifetimes of 1 ps or less, only limited by the NPL area.
However, for $L < 20$ nm excitons enter an intermediate confinement regime, hence departing from the 
quantum well behavior. 
 In heterostructured NPLs, 
 different response is observed for core/shell and core/crown configurations.
In the former, the strong vertical confinement limits separation of electrons and holes 
even for type-II band alignment. The exciton behavior is then similar to that in core-only NPL, 
albeit with weakened dielectric effects.
In the latter, 
 charge separation is also inefficient if band alignment is quasi-type-II (e.g. in CdSe/CdS), 
because electron-hole interaction drives both carriers into the core. However, it becomes very efficient
for type-II alignment, for which we predict exciton lifetimes reaching $\mu s$.
\end{abstract}

\maketitle

\section{Introduction}

Semiconductor NPLs are the colloidal analogous of epitaxial quantum wells.\cite{LhullierACR} 
Since the first CdSe NPLs were synthesized in 2008,\cite{IthurriaJACS} intensive research has revealed several
promising properties for optical and electronic applications, 
including flat surfaces with atomic monolayer control of thickness,\cite{IthurriaJACS,IthurriaNM,AchtsteinNL} 
fast and narrow fluorescent spectrum,\cite{IthurriaNM,AchtsteinNL,TessierACS12,NaeemPRB}
large exciton and biexciton binding energies,\cite{NaeemPRB,GrimNN}
high intrinsic absorption,\cite{AchtsteinJPCc} large two-photon-absorption cross section,\cite{ScottNL15,OlutasACS}
weak coupling of carriers to acoustic phonons,\cite{DongNL} fast and efficient fluorescence resonance energy transfer,\cite{RowlandNM}
suppressed multi-exciton recombination,\cite{KunnemanJPCL} and highly directionial emission.\cite{ScottXXX}

Despite clear similarities, the band structure of colloidal NPLs displays some remarkable differences with respect to epitaxial quantum wells.
The main physical factors inducing such differences are: (i) the intense interaction with organic ligands passivating their large surfaces,\cite{AchtsteinNL,EvenPCCP,BenchamekhPRB}
(ii) the presence of lateral confinement  and (iii) the possibility of growing heterostructures not only along the strong confinement direction, 
but also along the weakly confined one (so called core/crown NPLs).\cite{TessierNL13}

From a theoretical point of view, the effect of the organic environment has been addressed in different studies.
An important consequence is the build-up of significant dielectric mismatch between the inorganic NPL and its surroundings.
This has been estimated to enhance exciton binding up to few hundreds of meV.\cite{AchtsteinNL,BenchamekhPRB} 
Self-interaction of carriers with their image charges is also important. 
In NPLs it has been found to exceed the exciton binding energy enhancement, thus leading to a net blueshift of the emission energy.\cite{BenchamekhPRB,RodinaJETP,PolovitsynCM} 

The effect of lateral confinement has been less investigated. This is in spite of recent synthetic advances enabling the growth of
NPLs with lateral sizes down to few nm.\cite{BertrandCC}, and spectroscopic evidence that small area NPLs are less prone to non-radiative defects.\cite{OlutasACS}
Lateral confinement is also relevant in core/crown heterostructures, whose behavior differs qualitatively from their (sandwiched) core/shell counterparts.\cite{TessierNL13,LiACS}

A proper understanding of the influence of lateral confinement requires an accurate description of its interplay with electron-hole correlation.\cite{BryantPRB}
This is especially the case in NPLs, because Coulomb interactions are expected to be magnified by dielectric mismatch. 
Previous theoretical studies of NPLs with finite lateral size missed this requirement, as Coulomb interaction was either overlooked\cite{BoseJAP} 
or dealt with using a mean-field (Hartree) model.\cite{AchtsteinPRL} The latter has been tested in analyticaly solvable 2D systems and found 
to be insufficient, especially in the weak confinement regime.\cite{PlanellesXXX}

In the present work, we investigate exciton properties in core-only, core/shell and core/crown NPLs of CdSe.
The Schr\"odinger equation, including dielectric mismatch effects, is solved with a simple yet reliable variational wave function,\cite{PlanellesXXX}
which provides clear insight into the influence of correlation and lateral confinement.
We find large area CdSe NPLs lie in the weak confinement regime, where dielectric mismatch induces a nearly strict 2D behavior, 
but current available sizes also reach an intermediate confinement regime, where the lateral walls of the NPL increase binding energies,
emission energies and radiative recombination rates as compared to quantum wells.
Coulomb correlation is found to play a central role on all these properties. 
It is also responsible for the superradiance (or giant oscillator strength) observed in experiments,\cite{IthurriaNM,AchtsteinNL,NaeemPRB}
both in core-only and in heterostructured NPLs.

\section{Theoretical Model}

We calculate exciton states using the model proposed in Ref.~\onlinecite{PlanellesXXX} for quasi-2D structures.
We start with an effective mass Hamiltonian:
\begin{equation}
H = \sum_{i=e,h} \left( \frac{\vec{p}_\parallel\,^2}{2 m_{\parallel,i}} + \frac{p_z^2}{2 m_{z,i}} + V_i \right) + V_c(r_e,r_h) + E_{gap}^{\Gamma},
\label{eqH}
\end{equation}
\noindent where $e$ and $h$ stand for electron and hole, $m_{\parallel,i}$ is the in-plane mass of carrier $i$, $m_{z,i}$ that along the strongly confined direction ([001]), $\vec{p}_\parallel=(p_x,p_y)$ and $p_z$ the momentum operators, $V_i$ the single-particle potential, $V_c(r_e,r_h)$ the Coulomb interaction between electron and hole, and $E_{gap}^{\Gamma}$ the energy gap between conduction and valence band at the Gamma point.
The single particle potential is written as:
\begin{equation}
V_i = V_i^{pot} + V_i^{self}. 
\end{equation}
\noindent where $V_i^{pot}$ is the confining potential set by the band offset between the NPL core and its surrounding material, and $V_i^{self}$ the self-energy potential, resulting from the interaction of carriers with their image charges in the dielectric medium.
In core-only NPL, we take $V_i^{pot}=0$ inside the NPL and $V_i^{pot}=\infty$ outside.
For the self-energy potential, we consider image charges on top and bottom of the NPL are dominant, and take the corresponding quantum well expression:\cite{KumagaiPRB}
\begin{equation}
V_i^{self} = \sum_{n=\pm 1,\pm 2,\ldots} \frac{ q_n \, q_e^2}{2 \epsilon_1 \left[ z_i - (-1)^n z_i + n L_z \right]}.
\end{equation}
\noindent with $q_e$ being the electron charge, $q_n=((\epsilon_1-\epsilon_2)/(\epsilon_1 + \epsilon_2))^{|n|}$, 
$\epsilon_1$ ($\epsilon_2$) the dielectric constants inside (outside) the NPL and $L_z$ the thickness of the NPL.

For the electron-hole Coulomb interaction, we also take into account the influence of polarization due to the dielectric mismatch:\cite{KumagaiPRB}
\begin{equation}
V_c = \sum_{n=-\infty}^{\infty} \frac{ q_n \, q_e^2 } { \epsilon_1 \left[ (\vec{r}_{\parallel,e}-\vec{r}_{\parallel,h})^2  + \left[z_e - (-1)^n z_h + n\,L_z \right]^2 \right]^{1/2}}.
\label{eqVc}
\end{equation}

Hamiltonian (\ref{eqH}) is solved using a variational wave function of the form: 
\begin{equation}
\Psi = N \Phi_e \Phi_h \, e^{-\frac{\sqrt{(\vec{r}_{\parallel,e}-\vec{r}_{\parallel,h})^2}}{a_B^*}},
\label{eqPsi}
\end{equation}
\noindent where $\Phi_e$ and $\Phi_h$ are the electron and hole single-particle states, 
and the exponential term an in-plane Slater correlation factor, 
with $a_B^*$ the effective Bohr radius, which is optimized variationally. 
It is convenient to write $a_B^* = a_B^{3D}/(2\alpha)$, 
and let $\alpha$ be the actual variational parameter. 
Because we study quasi-2D systems, in most instances $\alpha$ will range between
$\alpha=0.5$, which corresponds to $a_B^*= a_B^{3D}$ (bulk Bohr radius) and $\alpha=1$, 
which corresponds to $a_B^*= a_B^{2D}$, which is the strictly 2D Bohr radius.

For the exciton ground state, band mixing is negligible,\cite{BoseJAP} so we use single-band 
particle-in-the-box states:
\begin{equation}
\Phi_i = \cos{k_x x_i}\, \cos{k_y y_i}\, \cos{k_z z_i}.
\label{eqPhi}
\end{equation}
 Here $k_x=\pi/L_x$, $k_y=\pi/L_y$ and $k_z=\pi/L_z$, with $L_x$, $L_y$ and $L_z$
being the dimensions of the NPL (see Fig.\ref{fig1}(a)). 
 Although $\Psi$ is best suited for square plates, with $L_x=L_y=L$ and $k_x=k_y=k$,
it can also be used to study rectangular NPLs as long as the short side is larger 
than the exciton Bohr radius, so that a single (in-plane isotropic) variational parameter 
is still a good approximation.

Notice in Eq.~(\ref{eqPsi}) that $\Psi$ correctly retrieves the expected behavior in the
limits of strong and weak lateral confinement. In the small $L$ limit, the exponential 
tends to unity and $\Psi \rightarrow N \Phi$, i.e. independent electron and hole states.
As $L$ becomes larger, $k$ decreases and $\Phi$ tends to the exciton center-of-mass function,
while the exponential describes the relative motion. In the limit, $k \rightarrow 0$, 
one obtains $\Psi \rightarrow N e^{-\sqrt{(\vec{r}_{\parallel,e}-\vec{r}_{\parallel,h})^2}/a_B^*}$,
i.e. the exciton relative motion with the lowest-energy center-of-mass state.

With our choice of $\Psi$, the minimization of $\langle \Psi | H | \Psi \rangle$ involves analytical 
and semi-analytical expressions, with integral multiplicity equal or lower than two for all terms.
Detailed derivation and expression can be found in Section 3 of Ref.~\onlinecite{PlanellesXXX}.
For illustration purposes and convenience for the analysis of results, here we only discuss the
normalization factor $N$. As can be inferred from Eq.~(\ref{eqPsi}), in cartesian coordinates
$N$ implies a six-dimensional integral. The integrals over $z_e$ and $z_h$ can be decoupled and
are analytical. In turn, the four-dimensional integral of in-plane coordinates can be reduced to
a two-dimensional one substituting $i_e$ and $i_h$ (with $i=x,y$) by $\xi_i^-=k_i i_e-k_i i_h$ and  
$\xi_i^+=k_i i_e+k_i i_h$. After trigonometric transformations, the integral over $\xi^+$ becomes
analytical and one obtains:\cite{PlanellesXXX,RomestainPRB}
\begin{multline}
N =  \frac{8 \pi^2}{L_x\,L_y\,L_z} \, \left[ \int_0^\pi d\xi_x^- \int_0^\pi d\xi_y^- \right. \\ 
\left. \,g(\xi_x^-)\,g(\xi_y^-)\, e^{-2a \sqrt{ (\xi_x^-/k_x)^2 + (\xi_y^-/k_y)^2 } } \right]^{-1/2},
\label{eqN}
\end{multline}
\noindent where
\begin{equation}
g(\xi^-_i) = (\pi - \xi^-_i)\,(2 + \cos{2 \xi^-_i}) + \frac{3}{2}\,\sin{2 \xi^-_i}.
\label{eqg}
\end{equation}

\noindent An alternative, analytical expression can be also obtained by integrating 
beyond the NPL lateral dimensions, over an infinite area:
\begin{multline}
N = \sqrt{\frac{2}{\pi}} \, \frac{8}{(L_x L_y)^{1/2} L_z}\,  \\
\left( \frac{1}{a^2} + \frac{a}{2}\,\sum_{i=x,y} \frac{1}{(a^2 + k_i^2)^{3/2}} + \frac{a}{4}\frac{1}{(a^2 + k_x^2 + k_y^2)^{3/2}} \right)^{-1/2}.
\label{eqNapp}
\end{multline}
\noindent where $a=1/a_B^*$. As we shall see, this approximation works well for large area NPLs,
and simplifies the analysis.

The exciton radiative rate in a dielectrically mismatched system is obtained as:\cite{LavallardAPPa}
\begin{equation}
\frac{1}{\tau} = \frac{n_{ext}}{n_{int}} \, \sum_{i=x,y,z} \tau_{int}^{-1}\, f_l(i)^2
\label{eq:tau}
\end{equation}
\noindent where $n_{ext}$ and $n_{int}$ are the refractive indexes outside and inside the NPL, respectively,
$f_{l}(i)$ is the local field factor along a given direction, which we calculate assuming ellipsoidal shape\cite{ShivolaJNM}, 
and $\tau_{int}^{-1}$ is the intrinsic radiative rate of excitons in the NPL, taken as homogeneous dielectric:
\begin{equation}
\frac{1}{\tau_{int}} = \frac{2 \, n_{int} \, q_e^2 \, E_{ph}\, E_{p}}{3 \pi \epsilon_0 \hbar^2 c^3 m_0}\, S_{eh}^2.
\label{eq:tauint}
\end{equation}
\noindent with $E_{ph}$ being the photon energy, $E_{p}$ the Kane matrix element, and 
$S_{eh}^2=| \langle \Psi | \delta_{r_e,r_h} | 0 \rangle |^2$ the electron-hole (envelope function) overlap.\\

Minor changes in the theoretical framework described above allow us to address different systems.
Excited $p$-exciton states can be modeled replacing e.g. $\cos (k_x x_i)$ by $\sin (2k_x x_i)$
in Eq.~(\ref{eqPhi}). Likewise, core/shell NPLs, heterostructured along the $z$ direction, can be 
modeled replacing $\cos (k_z z_i)$ by an appropriate numerical function. 
Type-I and quasi-type-II core/crown NPLs can be modeled by defining electron and hole cosine functions over 
different lengths, e.g. $L=L_{\mbox{{\tiny core}}}$ for holes and $L=L_{\mbox{\tiny crown}}$ for electrons.
It should be noted that every ratio of $L_{\mbox{\tiny crown}}/L_{\mbox{\tiny core}}$ implies different limits of integration
and different integrands in Eq.~(\ref{eqN}). In the Appendix we provide more details on this point. 
 Note also that, for simplicity, in heterostructures we assume the same effective mass and dielectric constant in the core and shell materials.

The reduced dimensionality of the numerical integrals involved in our model enables efficient computation. 
In the Supplemental Material we provide sample codes which can be used to calculate excitons in core-only, 
core/shell and core/crown NPLs (the latter for $L_{\mbox{\tiny crown}}=2L_{\mbox{\tiny core}}$).  
The codes are written in Mathematica. They have minimal memory requirements and a single run 
(for a given $\alpha$ parameter), with $V_i^{self}$ and $V_c$ sums including sufficient number 
of terms to achieve sub-meV converged results, lasts about a minute in an ordinary laptop (Intel Core i5-3317U processor).\cite{supmat}



\section{Results}

We investigate excitons in CdSe NPLs with and without shells.
CdSe material parameters are used: relative dielectric constant $\epsilon_1=10$, 
band gap at 300K $E_{gap}^{\Gamma}=1.658$ eV,\cite{Adachi_book} $E_p=16.5$ eV\cite{IthurriaNM} 
and non-parabolic effective masses. In the in-plane direction, the latter are taken from Ref.~\onlinecite{BenchamekhPRB}.  
In the $z$ direction, hole mass is taken as $m_{z,h}=0.9 m_0$,\cite{KimPRB} and the electron mass 
is fitted to reproduce the experimentally measured exciton emission energies,\cite{IthurriaNM} 
obtaining $m_{z,e}=0.4\,m_0$ ($m_{z,e}=0.3\,m_0$) for $4.5$ ($5.5$) monolayer (ML) NPLs.
For the ligands, we assume the dielectric constant to be $\epsilon_2=2$.\cite{AchtsteinNL}

\subsection{Core-only NPLs}
\label{s:co}

\begin{figure*}[htb]
\includegraphics[width=12.0cm]{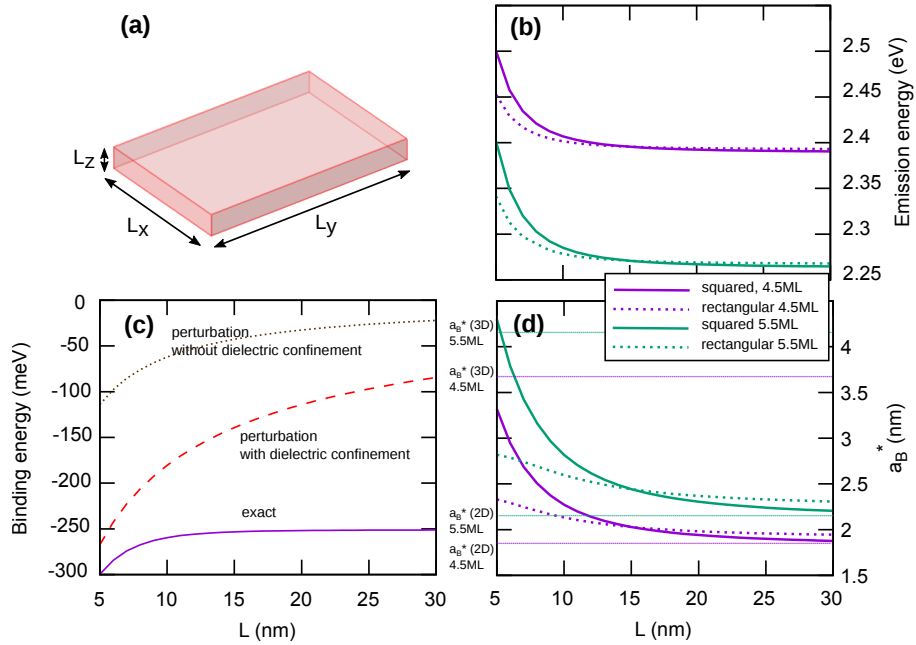}
\caption{(Color Online). (a) Schematic of the core-only NPL under study. 
(b) Exciton emission energy in square and rectangular CdSe NPLs with different size, shape and thickness.
Square NPLs have lateral dimensions $(L_x,L_y)=(L,L)$, and rectangular ones $(L,15)$ nm. 
Two thicknesses are considered, $L_z=1.4$ nm ($\sim 4.5$ ML) and $L_z=1.7$ nm ($\sim 5.5$ ML).
The effect of lateral confinement is felt under $L=15$ nm.
(c) Binding energy of the square NPL with $5.5$ ML thickness.
Solid line: fully correlated exciton.
Dashed line: perturbational estimate, neglecting correlation.
Dotted line: same but neglecting dielectric confinement.
(d) Effective Bohr radius for the NPLs of panel (a). Values lie between the 3D and 2D limit.
The latter is reached at $L \approx 20$ nm.
}
\label{fig1}
\end{figure*}

We start by investigating the effect of lateral confinement on cuboidal CdSe NPLs, 
like that represented in Fig.~\ref{fig1}(a). 
Fig.~\ref{fig1}(b) shows the exciton emission energy for square and rectangular NPLs
of different thickness ($\sim 4.5$ and $5.5$ MLs) as a function of the NPL side. 
In all cases, the exciton energy is in the infinite quantum well limit for sizes beyond 
$L \sim 15$ nm, but rapidly deviates for smaller sizes. Blueshifts of few to tens of meV 
can be expected in NPLs where at least one of the sides has $L<15$ nm, which is
consistent with recent experimental observations.\cite{BertrandCC} 
These shifts can be used for fine tuning of the emission energy, adding up to the 
step-like variations of energy imposed by the number of MLs in $z$. 
Yet, the shifts also introduce sensitivity to the exact lateral width,
which is less controllable than the thickness, and hence give rise to increased linewidths.
The latter effect could be detrimental e.g. for energy-transfer processes.\cite{RowlandNM}

The magnitude of the blueshift we estimate for $4.5$ ML NPLs with rectangular shape,
as the lateral size decreases down to 5 nm, is $\sim 50$ meV. This is in fact close to 
 (albeit slightly larger than) experimental values recently observed for similar dimensions 
in Ref.~\onlinecite{BertrandCC} ($\sim 25$ meV) and Ref.~\onlinecite{AchtsteinPRL} ($\sim 40$ meV).
Theoretical estimates using Hartree-renormalized multi-band k$\cdot$p theory yielded ~35 meV blueshift\cite{AchtsteinPRL},
in reasonable agreement with ours.
By contrast, calculations neglecting electron-hole interaction obtained (for square plates) blueshifts of 250-300 meV,\cite{BoseJAP}
about three times larger than our estimates for such geometry.
We conclude the balance between lateral confinement and  electron-hole interaction (at least to a Hartree level) is necessary to provide good estimates.

Lateral confinement also influences the binding energy. As shown in Fig.~\ref{fig1}(c)
--solid line--, for sizes $L<15$ nm the binding energy increases
because electron and hole are squeezed into a smaller area.
At this point, it is worth analyzing the origin of the binding energy. 
Dotted line in Fig.~\ref{fig1}(c) shows the exciton binding energy neglecting
dielectric confinement and electron-hole correlation. This is calculated with
a first-order perturbation description of Coulomb interaction, in the basis of
independent electron and hole states ($\alpha=0$). The resulting energy is relatively
weak and the effect of confinement is overestimated.
The inclusion of dielectric confinement (dashed line) drastically enhances 
binding energy,\cite{AchtsteinNL,BenchamekhPRB,RodinaJETP} 
but size dependence is still overestimated. 
Finally, the inclusion of full electron-hole correlation (solid line)
provides an additional enhancement of binding energies as well as an appropriate
description of the size independence as one approaches the quantum well limit,
which is set by the relative motion of electron and hole rather than 
center-of-mass confinement.
These results evidence that binding energy is largely determined by correlation,
which is particularly strong in NPLs owing to the dielectric mismatch.
 It is likely for this reason, that the binding energies we predict for $5.5$ ML NPLs 
($\sim 250$ meV) exceed previous calculations using perturbational or Hartree methods in 
NPLs with similar thickness and dielectric constants ($\sim 170$ meV).\cite{AchtsteinNL,BenchamekhPRB}
Ref.~\onlinecite{NaeemPRB} experiments have inferred a binding energy of $180$ meV
using multi-parameter fitting of the extinction spectrum. 
More direct and systematic measurements would be desirable to confirm the actual magnitude,
but the smaller value as compared to our prediction may be due to the fact that abrupt interface
models tend to overestimate dielectric confinement,\cite{EvenPCCP} so our model (and also 
 Refs.~\onlinecite{AchtsteinNL,BenchamekhPRB}) provide upperbound values.
The conclusion that correlation has a major effect in determining exciton binding energies is independent from this.

The effect of lateral confinement on the effective Bohr radius is plotted in
Fig.~\ref{fig1}(d). One can see for most $L$ values, $a_B^*$ lies in between the 2D
and the 3D values (dashed horizontal lines). Small Bohr radii, $a_B^* \rightarrow a_B^{2D}$, 
are prompted by the strength of correlation in these structures. However, for sizes
under $L=20$ nm the departure from the well behavior is apparent. 
Interestingly, the $a_B^{2D}$ limit is thickness-dependent. 
This is because the non-parabolicity of the bands translate into different in-plane masses 
for each thickness.\cite{BenchamekhPRB} Thinner NPLs have heavier masses, and hence 
$a_B^{2D}=\epsilon_1/2\mu_\parallel$ (atomic units used) decreases.

We conclude from Fig.~\ref{fig1} that CdSe NPLs with lateral sizes under 15-20 nm 
($\sim 6$ to $10$ times $a_B^*$)
are neither in the weak (quantum well) confinement regime, nor in the strong (quantum dot) one. Rather,
they lie in an intermediate regime where exciton energy and wave function are given by
the balance between electron-hole correlation and confinement.

The radiative lifetime of bright excitons in CdSe NLPs has been a subject of debate in the literature.
Photoluminescence experiments have measured lifetimes of 200-400 ps at low temperatures.
\cite{IthurriaNM,AchtsteinNL,TessierACS12,BiadalaNL} 
This is two orders of magnitude faster than in spherical CdSe nanocrystals, which has been interpreted 
as a manifestation of the giant oscillator strength effect (GOST) previously reported in epitaxial 
quantum wells.\cite{FeldmannPRL,HoursPRB} However, Naeem et al. pointed out that band edge lifetimes
would be shorter if one used resonant excitation. In fact, by means of four-wave mixing spectroscopy
at low temperature, they concluded
 in a NPL with $(L_x,L_y,L_z)=(30.8,7.1,1.67)$ nm $\tau$ is the ps range.\cite{NaeemPRB}
This would set CdSe NPLs among the fastest single-photon emitters, outperforming epitaxial structures.\cite{HoursPRB,TighenauPRL}

\begin{figure}[htb]
\includegraphics[width=8.6cm]{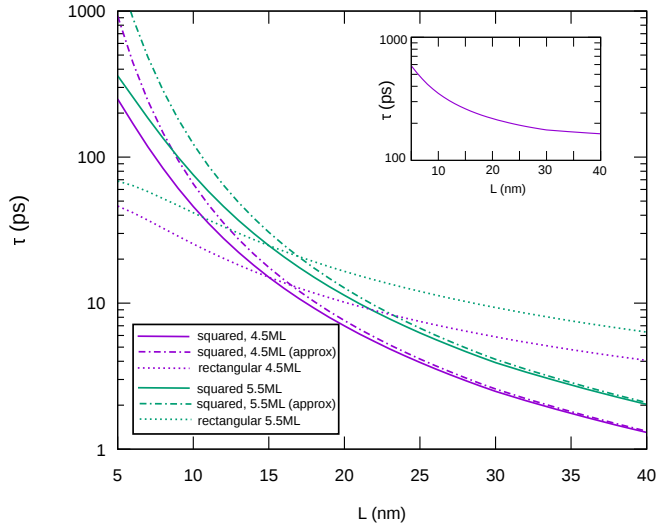}
\caption{(Color Online). Exciton radiative lifetime in CdSe NPLs with different lateral size, shape and thickness.
The lifetime is approximately proportional to $(a_B^*)^2 / (L_x\,L_y)$, as shown by dotted-dashed lines. 
Inset: lifetime of a $4.5$ ML square NPL neglecting electron-hole correlation. 
Notice they are orders-of-magnitude longer, and the size dependence is weaker.} 
\label{fig2}
\end{figure}

In Fig.~\ref{fig2} we provide theoretical assessment for this issue. We calculate the (ground state) bright exciton lifetime in square NPLs with lateral dimensions $(L,L)$ --solid lines-- and rectangular ones with $(L,15)$ nm --dotted lines--. A pronounced decrease with $L$ is observed in all instances, which is steeper for square NPLs. Our results support the observations of Ref.~\onlinecite{NaeemPRB} in that lifetimes of $\sim 1$ ps are feasible for typical NPL dimensions. In fact, extrapolating towards larger values of $L$ it is clear that sub-ps lifetimes are also feasible.\cite{coherent} 

The trends observed in Fig.~\ref{fig2} can be understood from the $S_{eh}^2$ factor in Eq.~(\ref{eq:tauint}).
From Eq.~(\ref{eqPsi}), the electron-hole overlap reads:
\begin{equation}
S_{eh}^2 = N^2\,
\left( \frac{L_x}{2} \right)^2 \,
\left( \frac{L_y}{2} \right)^2 \,
\left( \frac{L_z}{2} \right)^2. 
\label{eqSehexact}
\end{equation}
\noindent A simple analytical expression can be obtained replacing the normalization factor $N$ by 
Eq.~(\ref{eqNapp}) in the weak confinement regime ($L \gg a_B^*$), which implies $k \ll a$: 
\begin{equation}
S_{eh}^2 = \frac{8}{9\pi}\,L_x\,L_y\,\frac{1}{(a_B^*)^2}.
\label{eqSeh}
\end{equation}
\noindent From Fig.~\ref{fig1}(d), it follows for large NPLs we can further replace $a_B^*$ by 
the two-dimensional limit, $a_B^{2D}$. 

Lifetimes calculated with exact numerical $S_{eh}^2$, Eq.~(\ref{eqSehexact}), (full-lines in   Fig.~\ref{fig2}) and with analytic approximate  $S_{eh}^2$, Eq.~(\ref{eqSeh}), (dot-dashed lines) are quite alike. However, when the lateral confinement becomes significant, approximate analytical $S_{eh}^2$ gradually overestimates the exact numerical values, as shown in Fig.~\ref{fig2}.

Eq.~(\ref{eqSeh}) implies the exciton oscillator strength increases with the NPL area and the inverse of
the effective Bohr radius. In this sense, the behavior is consistent with the GOST theory proposed for 
quantum wells,\cite{FeldmannPRL} bound excitons,\cite{RorisonSM} and weakly confined parabolic 
quantum dots,\cite{TighenauPRL} except that the exciton coherence area here is set by the rigid walls. 
Note however that $a_B^*$ is particularly small in NPLs. Dielectric confinement squeezes the exciton
wave function,\cite{AchtsteinNL} and the 2D limit is already attained for lateral sizes of $L\approx 20$ nm, 
which are routinely synthesized. This confirms NPLs as particularly suitable nanostructures to exploit 
superradiance.

Fig.~\ref{fig2} also evidences that a small NPL thickness $L_z$, even if not explicit in Eq.~(\ref{eqSeh}),
leads to slightly shorter lifetimes. This is because of the heavier (non-parabolic) reduced mass 
of the exciton, which leads to smaller $a_B^*$.
It is also worth stressing that electron-hole correlation is at the heart of the GOST.
From Eq.~(\ref{eqPsi}) one can identify $a=1/a_B^*$ with the strength of the correlation. If we set $a=0$
(independent electron and hole, as in the strong confinement regime), the exciton lifetime is much longer
and only weakly dependent on the size (inset in Fig.~\ref{fig2}).\\

To test the robustness of the above results at finite temperatures, we have further calculated the electronic
properties of a few low-lying exciton states in square NPLs. 
The different symmetry of the states is reflected in the different quantum numbers of the 
particle-in-the-box function $\Phi_i(n_x, n_y, n_z)$. For the ground ($s$-like) exciton, 
$\Phi^{ss} = \Phi_e (1,1,1) \, \Phi_h (1,1,1)$, as in Eq.~(\ref{eqPhi}). 
Then we have optically dark excited states like 
$\Phi^{ps} = \Phi_e (2,1,1) \, \Phi_h (1,1,1)$ 
or 
$\Phi^{sp} = \Phi_e (1,1,1) \, \Phi_h (2,1,1)$.
Last, we consider the  optically active $p$-like exciton, 
$\Phi^{pp} = \Phi_e (2,1,1) \, \Phi_h (2,1,1)$.

\begin{figure}[htb]
\includegraphics[width=7.0cm]{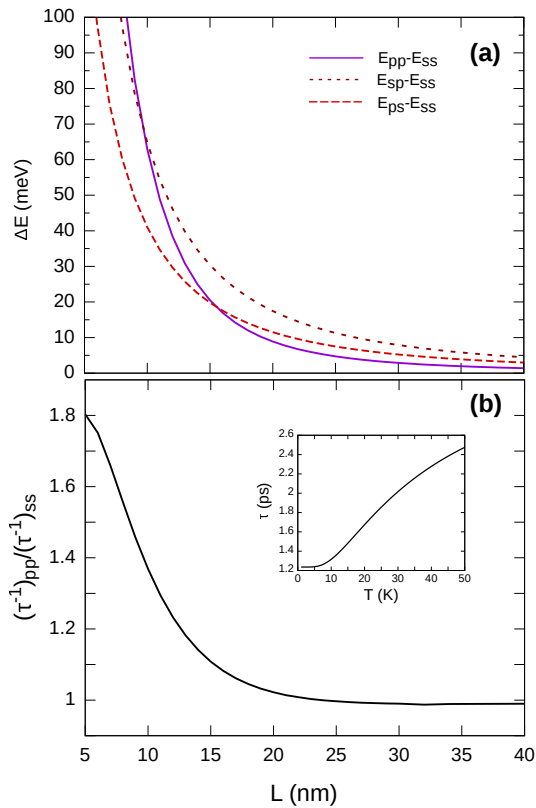}
\caption{(Color Online). (a) Energy splitting of excited exciton states ($\Phi^{ps},\,\Phi^{sp},\,\Phi^{pp}$) 
with respect to the ground state ($\Phi^{ss}$).  Solid (dashed) lines are used for optically bright (dark) states. 
(b) Ratio between radiative recombination rates of the ground ($\Phi^{ss}$) and first bright ($\Phi^{pp}$) exciton states.
Inset: thermally averaged recombination rate.  
The NPL dimensions are $(L_x,L_y,L_z)=(30,30,1.4)$ nm.}
\label{fig3}
\end{figure}

Fig.~\ref{fig3}(a) shows the energy splitting between the ground and the excited states,
as a function of lateral confinement. In all cases, the splitting decays rapidly
with $L$. For lateral sizes $L > 20$ nm, the splitting is of few meV only. 
 Fig.~\ref{fig3}(b) compares the radiative rate of the (optically active) $s$ and $p$ excitons.
The excited state recombines up to twice faster for small NPLs, but as the NPL side increases
soon both states present the same rate.
With the results above, we can calculate the thermally averaged radiative lifetime at finite
temperature $T$, which is shown in Fig.~\ref{fig3}(b) inset for a NPL with $L=30$ nm.
The lifetime increases with temperature due to the occupancy of dark states, but the ps regime is quite robust 
up to moderately high temperatures, in spite of the large area. 

It has been recently shown that the emission spectrum of CdSe NPLs often involves not only band edge exciton 
recombination, but also a second peak blueshifted by $20-40$ meV.\cite{AchtsteinPRL} 
Because the energy of the second peak was dependent on the lateral confinement, it was suggested it could arise 
from an excited $p$-exciton state, which would correspond to $\Phi^{pp}$ in our model. 
Unfortunately, our results cannot confirm this interpretation.
If the NPL has a lateral side $L>15$ nm, the predicted energy splittings are of few meV only. 
%
 On the other hand, the experiments showed the radiative recombination rate of the excited feature were much 
faster than those of the ground state. This is in contrast with Fig.~\ref{fig3}(b).  
Certainly, the results in Fig.~\ref{fig3} are of square NPLs and the experiment dealt with rectangular ones. 
However, we find similar behavior for rectangular NPLs, so this question remains open. 
Further experiments and related modeling is needed to rationalize the second peak in the emission spectrum.

\subsection{Heterostructured NPLs}
\label{s:cs}

CdSe NPLs can be sandwiched between layers of a different inorganic material for improved surface 
passivation.\cite{MahlerJACS}
Two important physical factors affect the exciton properties in that case.
First, the thickness of the shell tends to reduce the influence of the dielectric environment.\cite{PolovitsynCM}
Second, different core/shell band alignments can be used to concentrate or separate electrons 
from holes.  In this section we analyze these effects in more detail. 

\begin{figure}[htb]
\includegraphics[width=8.6cm]{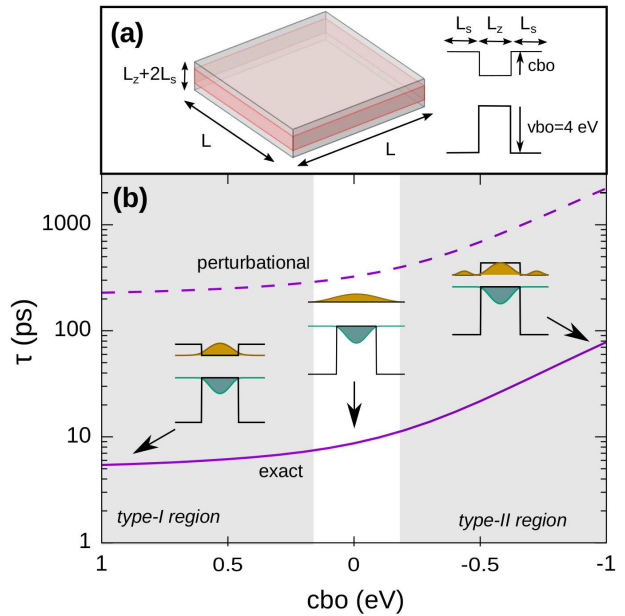}
\caption{(Color Online). (a) Schematic of a core/shell NPL and band-offset potential along $z$ direction.
(b) Exciton lifetime including (solid line) and excluding (dashed line) Coulomb correlation, as a function
of the $cbo$. The insets show the electron and hole charge density along the growth direction
of the NPL for the exact solution. As the electron and hole become separated, the exciton lifetime increases, 
but it remains superradiant.} 
\label{fig4}
\end{figure}

We consider square NPLs with $L=30$ nm and $L_z=1.4$ nm, surrounded by a shell with thickness $L_s=1$ nm  (see Fig.~\ref{fig4}(a)).
For simplicity, we take CdSe material parameters all over the structure, but manipulate the band-offsets to vary the
band alignment.  The valence band offset is fixed at $4$ eV, which forces the hole to stay in the core, but the conduction
band offset ($cbo$) is changed from positive (type-I band alignment) to zero (quasi-type-II) and to negative (type-II).
In this way, for $cbo > 0$ our system qualitatively resembles CdSe/(Cd)ZnS NPLs, for $cbo \approx 0$ it resembles CdSe/CdS NPLs,
and for $cbo < 0$ CdTe/CdSe NPLs.

In Fig.~\ref{fig4}(b) --solid line-- we plot the exciton lifetime as a function of $cbo$. As seen in the insets, 
moving towards type-II band alignment permits gradual electron delocalization into the shell. 
The decreasing electron-hole overlap $S_{eh}^2$ translates into extended lifetimes.
The extension turns out to be weak when changing from type-I to quasi-type-II, 
which is actually consistent with
experiments showing that CdSe/CdS NPLs display similar radiative rates to CdSe core only NPLs,\cite{KunnemanNL}
and it becomes more significant as the type-II band alignment is reached.
It is worth noting, however, that the strong confinement energy along the $z$ direction prevents complete charge separation 
(see right-most inset). As a result, even for type-II core/shell NPLs the exciton behavior resembles that in core-only NPLs,
with in-plane electron-hole correlation playing an important role.
In fact, for $cbo=-1$ eV, $\tau$ is one over one order of magnitude shorter than in a correlation-free exciton (dashed line).
In other words, superradiance is also important in type-II core/shell NPLs.

\begin{figure}[htb]
\includegraphics[width=8.0cm]{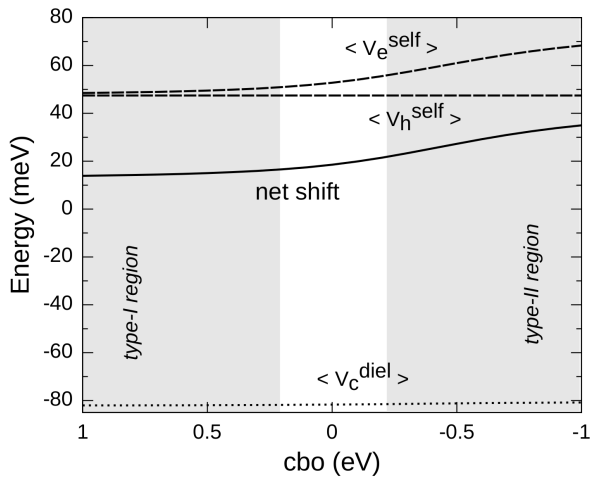}
\caption{Effect of electron-hole separation on dielectric confinement energetic terms.
Dashed lines: self-energy repulsion.
Dotted lines: electron-hole Coulomb attraction enhancement due to dielectric mismatch. 
Solid line: net effect on exciton emission energy.}
\label{fig5}
\end{figure}

 Core/shell band alignment also influences dielectric confinement. 
To illustrate this point, in Fig.~\ref{fig5} we plot how the different energetic terms induced
 by the dielectric mismatch vary with the $cbo$. 
We consider self-energy repulsion terms for electron and hole, $\langle  V_e^{self} \rangle$ and $\langle V_h^{self} \rangle$, 
the binding energy enhancement induced by the dielectric environment, 
$\langle V_c^{diel} \rangle =  \langle V_c \rangle - \langle V_{bare} \rangle$ --with $V_{bare}$ representing the bare Coulomb term, 
$n=0$ in Eq.~(\ref{eqVc})--,
and the net shift of emission energy, $\langle  V_e^{self} \rangle +\langle V_h^{self} \rangle + \langle V_c^{diel} \rangle$.
 In the type-I limit ($cbo=1$ eV), the exciton emission energy experiences a net blueshift of $15$ meV,
 which is very close to recent theoretical estimates for CdSe/ZnS NPLs.\cite{PolovitsynCM}
Such a blueshift, although with larger magnitude, was already reported in core-only CdSe NPLs,\cite{BenchamekhPRB,PolovitsynCM} 
and takes place because, in anisotropic objects like NPLs, the self-energy repulsion terms 
exceed the binding energy enhancement. 
The blueshift is less severe in type-I core/shell NPLs than in core-only ones because the shell
separates excitonic carriers from their image charges in the dielectric environment.
Yet, as one can see in Fig.~\ref{fig5}, the shift increases again as we move towards type-II band alignment. 
This is because the electron partly localizes into the shell, closer to its image charges, hence
increasing $\langle V_e^{self} \rangle$. Since $\langle V_c^{diel} \rangle$, in turn, is barely affected 
by the band offset (dotted line), the net result is a blueshift.

In addition to sandwiched core/shell NPLs, heterostructures with a CdSe core laterally surrounded by a CdX (X=S,Te) 
crown have been synthesized lately.\cite{TessierNL13,KelestemurJPCc,AntanovichNS} 
CdSe/CdS core/crown NPLs have quasi-type-II band alignment, and yet there is spectroscopic evidence that 
the band edge exciton is largely localized in the core, as in a type-I system.\cite{TessierNL13}
By contrast, in NPLs with type-II band alignment (e.g. CdSe/CdTe) a truly type-II exciton behavior has been found.\cite{KelestemurJPCc}

\begin{figure}[htb]
\includegraphics[width=8.0cm]{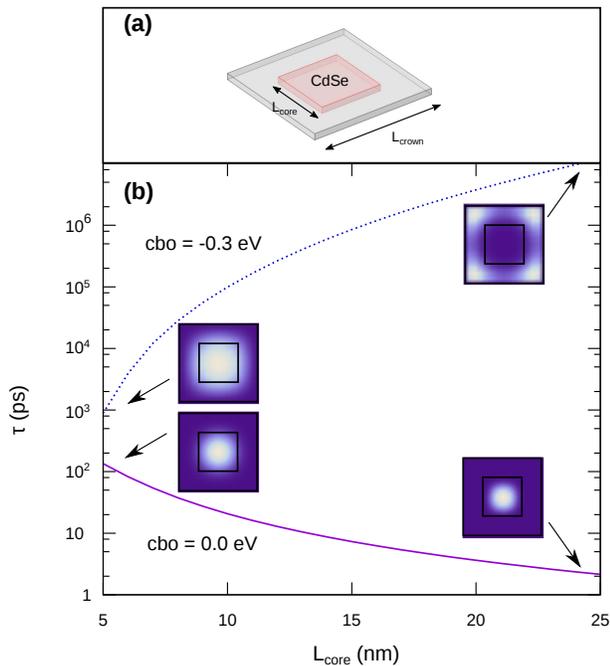}
\caption{(Color Online). (a) Schematic of a core/crown NPL.
(b) Exciton lifetime in quasi-type-II (solid line) and type-II (dotted line) core/crown NPLs
as a function of the core size. The insets show the electron charge density, with dark/light areas
indicating low/high density. The electron stays in the core for quasi-type-II band alignment, 
but it delocalizes into the shell for type-II ones.
 $L_{\mbox{{\tiny crown}}}= 2 L_{\mbox{{\tiny core}}}$.} 
\label{fig6}
\end{figure}

To offer theoretical insight into these systems, we calculate the exciton wave function and radiative
recombination time in core/crown NPLs with a CdSe core of lateral sides $L_{\mbox{{\tiny core}}}$ 
and a crown with $L_{\mbox{\tiny crown}}=2L_{\mbox{\tiny core}}$ (see Fig.~\ref{fig6}(a)). 
We first simulate a CdS shell by setting $cbo=0$ eV (perfect quasi-type-II band alignment) and
using:
\begin{equation}
\Phi_e = \cos{\frac{\pi x_e}{L_{\mbox{{\tiny crown}}}} }\,
         \cos{\frac{\pi y_e}{L_{\mbox{{\tiny crown}}}} }\,
         \cos{\frac{\pi z_e}{L_z}},
\end{equation}
\noindent and
\begin{equation}
\Phi_h = \cos{\frac{\pi x_h}{L_{\mbox{{\tiny core}}}}}\,
         \cos{\frac{\pi y_h}{L_{\mbox{{\tiny core}}}}}\,
         \cos{\frac{\pi z_h}{L_z}},
\end{equation}
\noindent in Eq.~(\ref{eqPsi}).
The resulting exciton function, as shown in the bottom insets of Fig.~\ref{fig6}(b), concentrates 
most of the electron charge density close to the core in spite of the band alignment, 
consistent with experiments.\cite{TessierNL13}
This is true even for the smallest cores we consider ($L=5$ nm), 
because the lateral confinement energy is much weaker than the exciton binding energy.
Consequently, the corresponding lifetime (solid line in the figure) behaves very much as that in core-only NPLs.

If we set a type-II band aligment instead, with $cbo=-0.3$ eV, the confinement and band offset potential energies 
can overcome the binding energy. Consequently,  as the crown size increases, the electron gradually localizes into the crown
(see top insets in Fig.~\ref{fig6}(b)).
Eventually, for large enough $L_{\mbox{\tiny crown}}$, the electron concentrates in the crown corners, 
where kinetic energy is minimized (upper-right inset). 
 The efficient charge separation translates into rapidly increasing lifetimes, 
which reach the $\mu s$ range for $L_{\mbox{{\tiny core}}} \gtrsim 20$ nm,
as shown by the dotted line in Fig.~\ref{fig6}(b).

The above results are in agreement with recent experiments using type-II (CdSe/CdTe and CdTe/CdSe) core/crown NPLs, 
which showed radiative lifetimes of up to $0.1$ $\mu s$ (for $16\times8$ nm$^2$ core), about two orders of magnitude
longer than in CdSe NPLs, evidencing charge separation in these structures.\cite{ScottPCCP} 
They are also consistent with experiments on quasi-type-II CdSe/CdS NPLs, where lifetimes have been 
found to resemble those in core-only structures,\cite{AntanovichNS}
which we interpret from the exciton localization in the core for $cbo \approx 0$ eV.
  
All in all, Fig.~\ref{fig6} shows that the behavior of core/crown NPLs is markedly 
different from that of core/shell ones.

\section{Conclusions}

We have shown that the physics of excitons confined in CdSe semiconductor 
homo- and heterostructured NPLs is largely governed by the interplay between the
strong electron-hole in-plane correlation, quantum confinement and dielectric confinement.
With these factors one can explain several experimental observations reported in the literature.
Simple and efficient (semi-analytical) codes are provided to calculate binding energies,
emission energies and electron-hole overlap integrals considering such effects.

Excitons in homostructured (core-only) CdSe NPLs are found to deviate from the quantum-well behavior 
for lateral sizes under $L=20$ nm (6-10 times the Bohr radius). An intermediate confinement regime
is then entered, where emission energy blueshifts and radiative lifetimes diverge from the 2D limit.
 Electron-hole correlations are shown to have a central role in determining the 
exciton properties. In particular, they are responsible for the GOST of 
laterally confined NPLs. Because Coulomb correlations are enhanced by dielectric confinement,
we argue colloidal NPLs are excellent material to exploit this optical feature.
 For realistic NPL sizes, we estimate band-edge exciton lifetimes of 1 ps or less, 
which supports related four-wave mixing measurements.\cite{NaeemPRB}
We also find that the lowest lying excited ($p$-)exciton state are a few meV higher in energy and exhibit similar lifetimes.

Excitons in sandwiched quasi-type-II or type-II core/shell NPLs show slightly 
longer lifetimes due to partial charge separation along the strong confinement direction. 
However, in-plane electron-hole correlation remains important and excitons are still superradiant. 
A qualitatively different behavior is observed in core/crown NPLs.
For quasi-type-II band alignement, such as CdSe/CdS structures, lateral confinement
energy cannot compete against Coulomb binding energy, and both carriers are confined inside the
core. This explains the core-only NPL behavior observed in experiments.\cite{TessierNL13}
Complete charge separation can be nonetheless achieved using type-II band alignment and large crowns.

\begin{acknowledgments}
We are grateful to Alexander Achtstein and Iwan Moreels for useful discussions.
We acknowledge support from MINECO project CTQ2014-60178-P and UJI project P1-1B2014-24.
\end{acknowledgments}

\appendix
\section{Reducing integral multiplicity in cuboids with asymmetric electron and hole confinement}
\label{s:app}

\begin{figure}[htb]
\includegraphics[width=6cm]{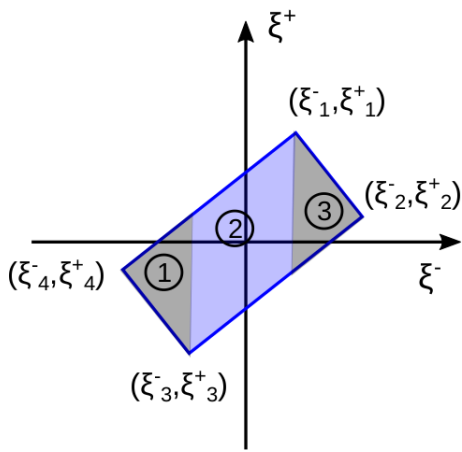}
\caption{Integration domain defined by the limits given in Eqs.~(\ref{eqdom1})-(\ref{eqdom4}). The circled numbers
label the three subdomains involved in Eq.~(\ref{eqN3}).}
\label{fig:dom}
\end{figure}

In cuboids with identical lateral confinment length for electron and hole, the Slater correlation
factor is known to permit reducing the in-plane integrals from four to two numerical integrals, 
as seen e.g. in Eq.~(\ref{eqN}).\cite{RomestainPRB}
Here we show such a reduction of integral multiplicity applies also to the 
 general case of a cuboid where electron and hole extend over different lengths. 
 For illustration purposes, we take a one-dimensional example 
(the extension to more dimensions is straightforward), with the wave function: 
\begin{equation}
\Psi = N \cos{k_e x_e} \, \cos{k_h x_h}\,e^{-a\sqrt{(x_e-x_h)^2}}, 
\end{equation}
\noindent where $k_e=\pi/L_e$ and $k_h=\pi/L_h$.  In a NPL with quasi-type-II band alignment, 
we could identify e.g.  $L_e=L_{\mbox{\tiny crown}}$ and $L_h=L_{\mbox{\tiny core}}$.
The normalization factor is obtained from a two-dimensional integral of the form:
\begin{equation}
N^2  = \int\limits_{\frac{-Le}{2}}^{\frac{Le}{2}} \int\limits_{\frac{-Lh}{2}}^{\frac{Lh}{2}} dx_e\,dx_h \cos{(k_e x_e)}^2 \, \cos{(k_h x_h)}^2\,f(|x_e,x_h|).
\end{equation}
\noindent where $f(|x_e,x_h|)$ is the correlation factor. 
We assume $p L_e = L$ and $q L_h = L$, $p,q \in N$, define the variables $\xi_e=k\,x_e$ and $\xi_h=k\,x_h$, with $k=\pi/L$, and rewrite the integral as:
\begin{equation}
N^2  = \frac{1}{k^2} \int\limits_{\frac{-\pi}{2p}}^{\frac{\pi}{2p}} \int\limits_{\frac{-\pi}{2q}}^{\frac{\pi}{2q}} d\xi_e\,d\xi_h \cos{(p \xi_e)}^2 \, \cos{(q \xi_h)}^2\,f(|\xi_e,\xi_h|).
\end{equation}
\noindent Next we switch to $\xi^+ = \xi_e + \xi_h$ and $\xi^- = \xi_e - \xi_h$ coordinates. 
The resulting integration domain becomes that in Fig.~\ref{fig:dom},
with the limit values:
\begin{align}
\label{eqdom1}
(\xi^-_1, \xi^+_1) &= (\frac{\pi}{2}\,\frac{p+q}{pq}, \frac{\pi}{2}\,\frac{q-p}{pq}), \\
(\xi^-_2, \xi^+_2) &= (\frac{\pi}{2}\,\frac{q-p}{pq}, \frac{\pi}{2}\,\frac{p+q}{pq}), \\
(\xi^-_3, \xi^+_3) &= (-\frac{\pi}{2}\,\frac{p+q}{pq}, -\frac{\pi}{2}\,\frac{q-p}{pq}), \\
(\xi^-_4, \xi^+_4) &= (-\frac{\pi}{2}\,\frac{q-p}{pq}, -\frac{\pi}{2}\,\frac{p+q}{pq})
\label{eqdom4}
\end{align}
\noindent and a $1/2$ Jacobian arises in the integrand.\\

\noindent The normalization factor can be now written as the sum of three integrals
over the independent subdomains of Fig.~\ref{fig:dom}:
\begin{equation}
N^2 = \frac{1}{2k^2} \sum_{i=1}^3 \iint_{\textcircled{i}} d\xi^+ d\xi^- I_{pq}(\xi^-,\xi^+)\,f(|\xi^-|), 
\label{eqN3}
\end{equation}
\noindent where $I_{pq} = \cos{(\frac{p}{2}\,(\xi^+ + \xi^-))}^2 \, \cos{(\frac{q}{2}\,(\xi^+ - \xi^-))}^2$.

For integer values of $p$ and $q$, the integral over $\xi^+$ is analytical within each subdomain: 
\begin{equation}
N^2 = \frac{1}{2k^2}\,\sum_{i=1}^3 \int_{\textcircled{i}} d\xi^- I_{pq}^{\textcircled{i}}(\xi^-)\,f(|\xi^-|).
\end{equation}
\noindent For example. for $p=1$ and $q=2$, one obtains:
\begin{eqnarray}
I_{12}^{\textcircled{1}} &=& \frac{1}{24} (12 \xi^- + 9 \pi + 8 \cos{2 \xi^-} - \sin{4 \xi^-}), \\
I_{12}^{\textcircled{2}} &=& \pi/4 + 2/3 \cos{2 \xi^-}, \\ 
I_{12}^{\textcircled{3}} &=& \frac{1}{24} (-12 \xi^- + 9 \pi + 8 \cos{2 \xi^-} + \sin{4 \xi^-})
\end{eqnarray}
\noindent which are the expressions we use for the core/crown NPL in section \ref{s:cs}. 

One can further check that for $p=q=1$, the integration domain becomes a symmetric rhomb, 
and the resulting integral $I^{pq}(\xi^-)$ can be identified with $g(\xi^-)$ in Eq.~(\ref{eqg}).


\begin{thebibliography}{40}
\bibitem{coherent} We are assuming the exciton coherence is set by the NPL area. For very large NPLs,
however, this may not be the case.
\end{thebibliography}
\end{document}